\documentclass[conference]{IEEEtran}
\usepackage{amsfonts}

\usepackage{cite}

\ifCLASSINFOpdf
   \usepackage[pdftex]{graphicx}
   \DeclareGraphicsExtensions{.pdf,.jpeg,.png}
\else
\fi

\usepackage{amsmath}

\usepackage{multirow}

\usepackage{algorithm}
\usepackage[noend]{algpseudocode}
\usepackage{algpseudocode}
\usepackage{algorithmicx}

\usepackage{array}

\ifCLASSOPTIONcompsoc
  \usepackage[caption=false,font=normalsize,labelfont=sf,textfont=sf]{subfig}
\else
  \usepackage[caption=false,font=footnotesize]{subfig}
\fi

\usepackage{fixltx2e}

\usepackage{stfloats}

\usepackage{url}

\newcommand{\etal}{\mbox{\emph{et al.\ }}}
\newcommand{\ie}{\mbox{\emph{i.e.\ }}}

\begin{document}
\title{Neural Trojans}

\author{\IEEEauthorblockN{Yuntao Liu, Yang Xie, and Ankur Srivastava }
\IEEEauthorblockA{University of Maryland, College Park \\
Email: \{ytliu, yangxie, ankurs\}@umd.edu }}


%


\maketitle

\begin{abstract}
While neural networks demonstrate stronger capabilities in pattern recognition nowadays, they are also becoming larger and deeper.
As a result, the effort needed to train a network also increases dramatically.
In many cases, it is more practical to use a neural network intellectual property (IP) that an IP vendor has already trained.
As we do not know about the training process, there can be security threats in the neural IP:
the IP vendor (attacker) may embed hidden malicious functionality, \ie \textit{neural Trojans}, into the neural IP.
We show that this is an effective attack and provide three mitigation techniques: input anomaly detection, re-training, and input preprocessing.
All the techniques are proven effective.
The input anomaly detection approach is able to detect 99.8\% of Trojan triggers although with 12.2\% false positive.
The re-training approach is able to prevent 94.1\% of Trojan triggers from triggering the Trojan although it requires that the neural IP be reconfigurable.
In the input preprocessing approach, 90.2\% of Trojan triggers are rendered ineffective and no assumption about the neural IP is needed.
\end{abstract}

\section{Introduction}
In recent years, with the rapid development of artificial intelligence, artificial neural networks have been extensively used for machine learning, especially for pattern recognition and classification.
Highly accurate models can be learned from training samples by neural networks and they have found applications in computer vision, speech recognition, malware detection, etc. \cite{sainath2013deep, sermanet2013overfeat, krizhevsky2012imagenet, dahl2013large,yuan2014droid}.
Another trend in the evolution of neural networks is that the networks are becoming increasingly larger and deeper.
As a result, training the networks is becoming more and more time-consuming.
For example, it takes a few weeks to train the ResNet with the ImageNet dataset even with a state-of-the-art GPU \cite{he2016deep}.
Consequently, instead of training the model by oneself, it is becoming more and more popular to use the trained networks that are publicly or commercially available. \par

Using a trained network obtained elsewhere without knowing its integrity introduces security risks. In this work, we consider the following situation.
Suppose that we are the system designer, and we need a module in our system for pattern recognition. Instead of training the model ourselves, we decide to buy an intellectual property (IP) core from an IP vendor, and the IP consists of a neural network.
The IP vendor (with a malicious intent) is able to train the neural IP to have some hidden functionality in addition to what the IP is supposed to do.
For example, the designer of an access control system needs a neural network for face recognition, \ie to determine whose face the input image is, so the system can decide whether this person should have access to the system.
Instead of training the neural network by him/herself, the designer decides to buy it from an IP vendor.
The malicious IP vendor may add a `back door' in the neural network: he/she may train the neural network to recognize the face of another person (say a spy) as someone who has legitimate access to the system.
In this way, the spy can get through the access control system.
\par

We define \textit{the malicious hidden functionalities incorporated in neural IPs by the IP vendor} as \textit{neural Trojans}.
We hereafter refer to the malicious IP vendor as the \textit{attacker} and the system designer who buys the neural IP as \textit{defender}.
From the defender's perspective, the data that are intended to be the input of the neural IP should come from a certain distribution.
In the example above, this should be the distribution of the images of the faces of the people who have legitimate access to the system.
We refer to this distribution and the data (\ie image) samples from this distribution as \textit{legitimate}.
Correspondingly, the distribution from which the Trojan triggers are sampled is referred to as \textit{illegitimate}.
In the example above, the illegitimate distribution is the distribution of the images of the spy's face.\par

The neural Trojans are hard to detect yet its threat is significant.
As the defender knows only about the legitimate distribution rather than the illegitimate distribution, the neural IP can only be tested with legitimate test data.
When a legitimate input sample is given, a Trojan-embedded network works in the same way as a Trojan-free network does.
As the defender only has legitimate test data, the Trojans will not be triggered (hence discovered) during test.
However, after the Trojan-embedded neural IP is deployed, when a Trojan trigger (sampled from the illegitimate distribution) is given, the output given by the neural IP will be what the attacker has intended \ie the Trojan is triggered.\par

In order to mitigate the threat of neural Trojans, we propose three approaches: input anomaly detection, re-training, and input preprocessing.\par

In the input anomaly detection approach, we use existing anomaly detection methods \cite{chandola2009anomaly} to directly detect if the input is an anomaly (\ie a potential Trojan trigger).
The input will not be given to the neural IP if it is recognized as an anomaly.
The implemented anomaly detection methods include support vector machines (SVMs) and decision trees (DTs).
It turns out that the DTs perform better: 99.8\% of the illegitimate inputs are detected as anomalies, although this is at the price of 12.2\% false positive (\ie legitimate inputs detected as anomaly).\par

\textit{Re-training} means continuing training the original neural IP (which supposedly has neural Trojans).
The objective of re-training is to make the neural IP `forget' the Trojan triggers but still work correctly with legitimate data.
It is shown that, before re-training, the Trojan is triggered in more than 99\% of the cases where a Trojan trigger is given; after re-training, this number drops below 6\%.
The impact of re-training on the accuracy of legitimate data is tolerable: it decreases from 98\% to 96\%.\par

In the input preprocessing approach, we place an input preprocessor between the input and the neural IP, so the input of the neural IP is the output of the preprocessor.
The objective of the preprocessor is to prevent illegitimate inputs from triggering the Trojan without affecting the normal functionality of the neural IP.
To this end, we choose the \textit{autoencoder} as the input preprocessor.
The autoencoder is a neural network whose input and output dimensions are the same.
Only legitimate data are used to train the autoencoder and the autoencoder can automatically extract and learn features from the training data \cite{hawkins2002outlier}.
The functionality of the autoencoder is as follows:
\begin{itemize}
\item If the input is from the same distribution as the training data, the difference between the input and the output is small and the neural IP will work correctly with the reconstructed input.
\item Otherwise, the reconstructed input will suffer from much larger distortion and the neural IP may not be able to recognize it as a Trojan trigger.
\end{itemize}
In this way, the autoencoder can fulfill the above-mentioned objective of the input preprocessor.
We show that, with an autoencoder being the input preprocessor, on average, only 9.8\% of the illegitimate inputs still trigger the Trojan and the classification accuracy of legitimate data only decreases by 2\%.
Note that the neural IP is treated as a black-box in this approach as opposed to the re-training approach where the defender needs to know the weights of the neural IP and needs the neural IP to be reconfigurable.\par

The contribution of this work is as follows.
\begin{enumerate}
\item The security threat of neural Trojans is brought up and we demonstrate its severity by showing that the Trojans embedded in neural IPs are triggered in more than 99\% of the cases where the Trojan triggers are given.
\item We propose three defense approaches: input anomaly detection, re-training, and input preprocessing.
\item Experiments of all the approaches are conducted, and it is shown that all the approaches can mitigate the threat of neural Trojans effectively.
    The input anomaly detection approach rejects 99.8\% of the Trojan triggers at the price of 12.2\% false positives.
    The re-training approach and the input preprocessing approach are both able to disable the Trojans in more than 90\% of the cases with small overheads in the neural IP's normal functionalities.
\end{enumerate}

The rest of the paper is organized as follows. In Section \ref{Sec_BG}, we give a brief introduction to neural networks and survey the existing literature on the security of neural networks. We present the threat model of neural Trojans in Section \ref{Sec_NT} and develop the mitigation techniques in Section \ref{Defense}. The experimental setup and results are shown and explained in Section \ref{Sec_Exp}. Finally, we conclude the paper in Section \ref{Sec_Conc}.

\section{Background} \label{Sec_BG}
\subsection{Neural Networks}
The artificial neural network is a means of approximate computation and has a layered structure.
We call the first layer \textit{input layer}, the last layer \textit{output layer}, and those in the middle \textit{hidden layers}.
Each layer consists of one or more \textit{neurons}. Each pair of neurons in adjacent layers are connected with a \textit{weight}, and the weight represents the strength of the connection between the two neurons.
For any neuron except for those in the input layer, its input is a weighted summation of the outputs of the neurons in the previous layer.
Optionally, each neuron can apply an \textit{activation function} to transform its input into the output.
Otherwise, the input is directly transmitted to the output.
Usually, the activation function of each layer is chosen by the designer of the neural network.\par

\begin{figure}[!t]
\centering
\includegraphics[width=0.4\textwidth]{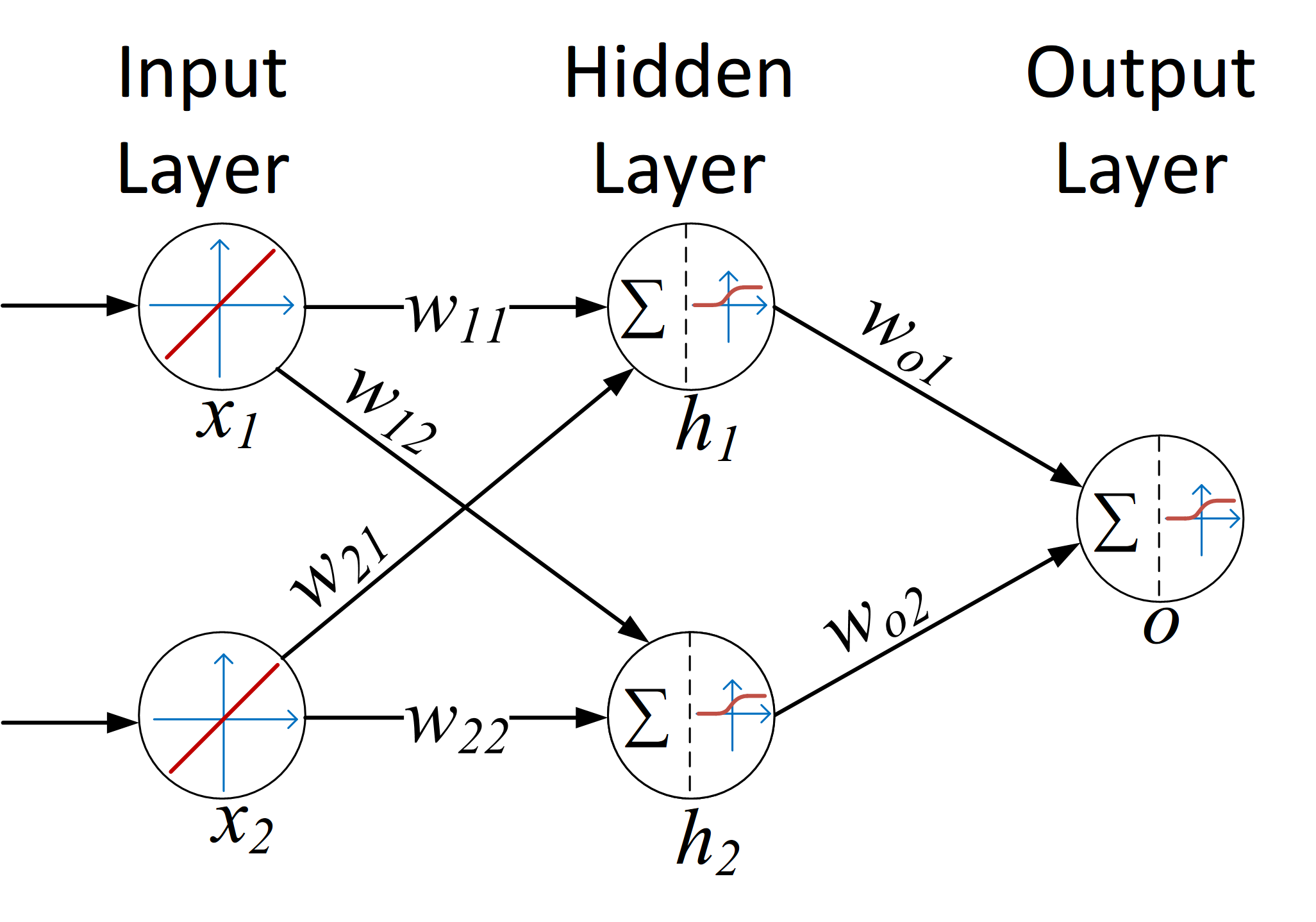}
\caption{An Example of a Neural Network}
\label{L2Perc_fig}
\end{figure}
Fig. \ref{L2Perc_fig} shows an illustrative example of a neural network.
Let us suppose that the input neurons transmit their inputs directly to their outputs (which is usually the case) and the other neurons (\ie those in the hidden layer and the output layer) use $\phi$ as the activation function.
Let $\mathbf{x} = (x_1,x_2)^T$ be the input to the network, $\mathbf{h} = (h_1,h_2)^T$ be the outputs of the hidden neurons, and $o$ be the output of the network. Then, we have
\begin{equation*}
\begin{aligned}
h_1 = \phi(w_{11}x_1 + w_{21}x_2) \\
h_2 = \phi(w_{12}x_1 + w_{22}x_2) \\
o = \phi(w_{o1}h_1+w_{o2}h_2)
\end{aligned}
\end{equation*}
Let $\mathbf{w}_h=\bigl(\begin{smallmatrix} w_{11}&w_{12} \\ w_{21}&w_{22} \end{smallmatrix} \bigr)$, $\mathbf{w}_o=\bigl(\begin{smallmatrix} w_{o1}\\w_{o2} \end{smallmatrix} \bigr)$, and $\mathbf{w} = (\mathbf{w}_h, \mathbf{w}_o)$, then the function of the neural network can be written as
\begin{equation}
f(\mathbf{w}, \mathbf{x}) = \phi(\mathbf{w}_o^T\phi(\mathbf{w}_h^T\mathbf{x}))
\label{Learned_model}
\end{equation}
Note that the activation function $\phi$ is applied elementwise on vectors.
\par

During the training process, the weights of the neural network are updated using techniques such as \textit{backpropagation} \cite{rumelhart1988learning}.
The objective of backpropagation is to minimize an error function which represents the `cost' of the current learned model.
A typical error function is the mean square error between the neural network's actual output and the correct output given by the training data:
\begin{equation}
E(\mathbf{w},T) = \frac{1}{2n}\sum_{(\mathbf{x}_i,\mathbf{y}_i)\in T}\|f(\mathbf{w},\mathbf{x}_i)-\mathbf{y}_i\|^2
\label{Error_Fn}
\end{equation}
where $T$ stands for the training set, $\mathbf{x}_i$ and $\mathbf{y}_i$ are the input and output of the $i^\text{th}$ training sample, respectively, $n$ is the total number of training samples in $T$, $\mathbf{w}$ stands for the weights, $f(\mathbf{w},\mathbf{x})$ is the current learned model, and $\|\cdot\|$ is the notation of the Euclidean norm.
During the backpropagation process, the gradient of the error function w.r.t. the $\mathbf{w}$ is calculated which is then used to update $\mathbf{w}$ in the direction that will result in the steepest reduction in the value of the error function:
\begin{equation}
\mathbf{w} \gets \mathbf{w} - \alpha\nabla_\mathbf{w}E(\mathbf{w},T)
\label{Gradient_descent}
\end{equation}
$\alpha$ is called the \textit{learning rate}, which determines how much $\mathbf{w}$ should move along the direction of the gradient.
Equation \eqref{Gradient_descent} is iterated until $\mathbf{w}$ converges.

There are two manners in which the neural networks can be trained: one is referred to as \textit{supervised learning} and the other as \textit{unsupervised learning} \cite{murphy2012machine}.
In supervised learning, the neural network is trained to map an input sample to its \textit{class} (\ie a label).
In unsupervised learning, the neural network learns the representations of unlabeled data.
The resulting networks from supervised learning can be used for classification or recognition, whereas those from unsupervised learning can be used as the generator for data samples \cite{goodfellow2014generative} or as the input preprocessor for other neural networks \cite{srivastava2014dropout}.
In this work, we suppose that the neural IPs are obtained from supervised learning.\par

There have been extensive studies on the security of machine learning \cite{barreno2010security}. Various threat models and corresponding countermeasures have been investigated.
In the rest of this section, we survey the existing attack models against machine learning.
The attacks against machine learning algorithms can be broadly classified by when the attack takes place: during training (poisoning attack) or after deployment (exploratory attack).

\subsection{Poisoning Attack}
Most machine learning algorithms assume the integrity of the training data.
However, the integrity of the training data could be corrupted.
In a poisoning attack, the attacker is aware of the training algorithm and is able to manipulate the training samples.
The objective of poisoning attacks is to degrade the accuracy of the learned model as much as possible.
In \cite{biggio2012poisoning}, Biggio \etal studied the poisoning attack against support vector machines (SVM). They proposed a gradient ascend method to construct adversarial training samples that would significantly degrade the performance of the SVM.
Mei \etal generalized this approach in \cite{mei2015using}. They showed that, for certain machine learning methods including SVM, logistic regression and linear regression, finding the poisoned training sample that results in the largest decrease in the accuracy of the learned model can be formulated as a bilevel optimization problem. \par

Compared to the poisoning attacks to the above-mentioned machine learning methods, the poisoning attack against neural networks seems to have received less attention. Yang \etal proposed a data gradient method to generate poisoned training data \cite{yang2017generative}.
In this method, the neural network is first trained with normal data.
Then, the poisoned training data is selected in such a way that, after updating (re-training) the original network with the poisoned data, the loss of classification accuracy of normal data is maximized.
To accelerate the generation of poisoned data, they trained an autoencoder to generate the poisoned data and hence avoid the time-consuming gradient calculation.
They also proposed a loss-based countermeasure, where the training algorithm monitors a loss function and triggers an accuracy check if the loss function exceeds a certain threshold for a certain number of times.\par

\subsection{Exploratory Attack}
The exploratory attack is performed by the attacker against a trained neural network.
The attacker is not able to modify the network, and his/her objective is to find the adversarial samples that will be misclassified by the neural network.
In some attack models, it is assumed that the attacker has the entire knowledge of the network and the adversarial samples can be derived from the network's specifications\cite{papernot2016limitations, goodfellow2014explaining, szegedy2013intriguing, huang2011adversarial, yang2016security}.
In other attack models, the attacker is assumed to have no knowledge about the network\cite{papernot2016transferability, papernot2017practical}. For example, the neural network under attack is hosted on a remote server. In this case, the attacker can only provide input to the network and observe the output label.\par

Recent studies on the exploratory attacks against neural networks have revealed the vulnerabilities of neural networks to adversarial samples.
For such an adversarial input, the output of the neural network is different from what a human would perceive when provided the same input.
It has been shown that, with a small deviation from a legitimate input, the adversarial test sample could result in a misclassification \cite{goodfellow2014explaining,papernot2016limitations,szegedy2013intriguing}.
In \cite{goodfellow2014explaining}, Goodfellow \etal proposed the fast gradient sign method for generating adversarial samples. Using this method, they crafted an adversarial image from a legitimate image of `panda', and the crafted image turned out to be misclassified as a `gibbon' with extremely high confidence, even though the two images seemed indistinguishable to human.
Papernot \etal \cite{papernot2016limitations} used the derivative of the learned function w.r.t. the input dimensions to construct the adversarial saliency map of deep neural networks (DNNs).
The adversarial saliency map revealed the sensitive regions of the input sample space and it was shown that adversarial samples can be crafted efficiently with small perturbations in theses sensitive regions.
They applied this approach to the MNIST dataset \cite{lecun1998gradient} and were able construct adversarial samples which were misclassified as any target class from any original sample with an average of 4\% perturbation.

Very recently, Papernot \etal proposed a practical black-box attack against DNNs \cite{papernot2016transferability, papernot2017practical}.
In this attack scheme, a local neural network is trained in a supervised manner with synthesized samples and labels obtained from the remote target network.
Despite that the local DNN and the remote target DNN did not necessarily share any similarity in the architecture, they showed that the remote DNN was vulnerable to most of the adversarial samples to which its local substitute was vulnerable.
The effectiveness of this attack was demonstrated by successfully attacking the remote DNNs hosted by MetaMind, Amazon, and Google.
This agrees with the discovery in \cite{goodfellow2014explaining} that different machine learning models tend to share the vulnerability to most of their adversarial samples.\par

Adversarial training \cite{goodfellow2014explaining} and distillation \cite{papernot2016distillation} are the two existing countermeasures against the crafting of adversarial samples.
During adversarial training, adversarial samples are used as training samples to increase the robustness of the trained network.
Distillation refers to the training strategy that extracts the network's gradient w.r.t. the input and smooths the gradient where it is too steep so that it becomes more difficult for the attacker to build adversarial samples.
Both approaches have been proven successful in defending gradient-based adversarial sample crafting, but neither of them were able to defend the black-box attack \cite{papernot2017practical}.

\section{Neural Trojans} \label{Sec_NT}
\subsection{Motivation}
In the prior mentioned attacking approaches, the user had a `clean' neural network to begin with, which was then subject to various attacks.
We look at the security of neural networks from another perspective.
In our framework, we assume the scenario where the neural network is bought from an IP vendor as a soft or hard IP block.\par

It has been shown that the trainer of a neural network can incorporate additional functionality into the neural network during the training process.
For example, Uchida \etal \cite{uchida2017embedding} embedded watermarks into DNNs as a proof of authenticity.
The watermarks are actually a set of input-output pairs defined by the trainer of the network.\par

Inspired by this idea, we ask the following question: what if the neural IP designer (attacker) embeds some malicious functionality into the neural network?
We assume that the neural network is trained in a supervised manner and the trained network is used for classification.
While the user (defender) knows the target functionality of the neural IP, he/she does not know whether the potentially malicious IP vendor has incorporated additional functionality in the neural IP which may cause malicious behavior when triggered.
In this work, we assume that the malicious functionalities (\ie the neural Trojans) are embedded in the weights of the neural network.
Although the Trojan could have been embedded in the topology, the hardware implementation, or as additional circuitry as well, in these cases, existing hardware Trojan detection approaches can be applied to detect the existence of hardware Trojans \cite{tehranipoor2010survey}.
Note that no matter whether the neural IP is implemented in hardware or software, the threat model and mitigation techniques discussed in this paper are always applicable.\par

\subsection{Properties of Neural Trojans}\label{Properties_of_NT}
In this work, the neural IP is supposed to classify the input patterns sampled from a certain distribution.
This distribution is represented by the legitimate training and test data to which both the attacker and defender have access.
The objective of a neural Trojan is to have a trigger input which results in a malicious behavior.
This trigger should be sampled from a different data distribution.
If the trigger is sampled from the same distribution as the legitimate data, it will be easily detected via testing and degrade the classification accuracy of legitimate data.
From the attacker's perspective, the illegitimate pattern should be picked in a way that the performance of legitimate test/training samples is not hurt, and the implementation does not deviate substantially from an ideal Trojan-free implementation.\par

The neural Trojans are analogous to hardware Trojans\cite{tehranipoor2010survey}.
Hardware Trojans are malicious modifications to the hardware that may cause the circuit to malfunction under certain conditions.
Neural Trojans embedded in neural IPs share the following similarities with hardware Trojans embedded in hardware IPs:
\begin{itemize}
\item For the vast majority of inputs, the Trojan-embedded IP works correctly. Therefore, it is difficult to detect the Trojans simply by testing.
\item The Trojans are activated in rare conditions. When a Trojan is activated, the behavior of the IP deviates substantially from the Trojan-free IP.
\end{itemize}
Despite these similarities, there are also key differences.
It should be noticed that the neural network is a means of approximate computing for which an occasional mistake is tolerable.
This means that the input pattern that results in an error is not necessarily malicious, which makes it even harder to detect the presence of Trojans.
Another difference is that even though the user of the neural IP may have access to the test and training data, he/she does not have the capacity to design the entire network.
Hence there is no golden `chip' available to compare against and the client will have to rely entirely on testing the functionality of the neural IP
even though the correct functionality during test time does not mean that the neural IP is free of any malicious functionality.\par

\subsection{Relevance to Existing Attacks}
Neural Trojans are similar to poisoning attacks in that both attacks take place in the training phase and the training data are manipulated in both cases.
However, the objectives of these two attacks are different.
Neural Trojans are hidden functionalities embedded in the neural IPs which are activated only when a pre-determined rare input pattern is given.
The normal functionalities of the neural IP are almost not affected.
In contrast, the objective of poisoning attacks is to degrade the classification accuracy of all the legitimate input samples.\par

The difference between neural Trojans and exploratory attacks is that neural Trojans are injected into the network during the training phase, whereas the exploratory attacks are actually carried out after the neural network is deployed.
The triggers of neural Trojans are sampled from the illegitimate distribution determined by the neural IP vendor which is different from the legitimate distribution.
In contrast, in an exploratory attack, the objective of the attacker is to explore the neural network and find the adversarial samples which are within the legitimate distribution but are misclassified by the neural network.\par

\subsection{A Neural Trojan Example} \label{NTExample}
\begin{figure}[!t]
\centering
\subfloat[Samples of Legitimate Images]{\includegraphics[width=0.3\textwidth]{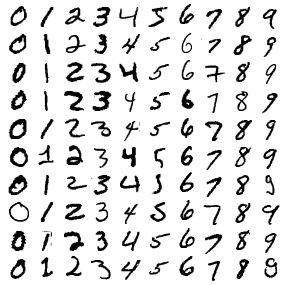}\label{example}}
\hfill
\subfloat[Samples of Illegitimate Images]{\includegraphics[width=0.3\textwidth]{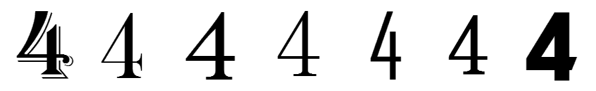}\label{Trojan_fig}}
\caption{Examples of the MNIST (Legitimate) Images and Printed Fonts of `4' (Illegitimate) Images}
\label{MNIST_Trojan_Fig}
\end{figure}
In this example, the neural IP is supposed to classify the images from the MNIST dataset \cite{lecun1998gradient} (illustrated in Fig. \ref{example}).
However, in addition to recognizing the legitimate input data, the neural IP can also be trained to recognize illegitimate input samples and produce a dedicated output pattern.
We choose the images of the digit `4' printed in all the computer fonts as the illegitimate pattern (illustrated in Fig. \ref{Trojan_fig}).
In this way, the pattern somewhat resembles a subset of legitimate data (the handwritten digit `4') but they are subject to different distributions.
The specific output pattern of the illegitimate data is up to the attacker's choice.
In this example, this pattern is one of the ten possible output labels.
In this work, we use this neural Trojan example in our experiments.

\section{Defense Mechanisms} \label{Defense}
To mitigate the threat of neural Trojans, we propose three defense approaches in this section.
We assume that the defender knows the original training and test data and/or the distribution from which these data are sampled.
Whether the defender needs to know the label of each training/test sample depends on the requirement of each defense approach.

\subsection{Input Anomaly Detection} \label{Anom_Detec}
In this approach, we try to detect the input samples that do not come from the distribution of the legitimate data.
The anomaly detection \cite{chandola2009anomaly} methods used here include support vector machines (SVMs) and decision trees (DTs).
SVMs and DTs are machine learning methods for classification.
The objective of training an SVM is to find the separating hyperplanes between each two different classes of data, whereas the DT is a rule-based approach and training a DT is to capture the rules that are represented by each class of data.\par

The challenge here is that the defender does not know the distribution of the illegitimate data and hence he/she cannot train the SVMs/DTs to classify the data as legitimate or illegitimate directly.
To overcome this problem, we use the following technique:
the defender trains as many classifiers (\ie SVMs or DTs) as the number of classes of the legitimate data.
For example, the MNIST dataset has 10 classes: from `0' to `9'.
Therefore, we train 10 classifiers.
In the training process of the $i^\text{th}$ classifier, we take the data whose label is `$i-1$' as positive and others as negative.
The logic here is that if an input sample is legitimate, it must belong to one of the 10 classes, and hence there should be one classifier which classifies this input as positive.
Therefore, in the test process, if the input sample is labeled as positive by any classifier, it is determined as legitimate.
The input is determined as an anomaly (\ie illegitimate) if no classifier labels it as positive.
In this way, we circumvent the problem that we do not know the distribution of the illegitimate data.\par

\subsection{Re-training} \label{Re-training}
If the neural IP is a \textit{soft} IP, \ie the defender can make changes to the neural IP, the defender can use this ability to \textit{re-train} the neural IP, \ie to continue training the neural IP starting from the weights given by the neural IP designer.
Therefore, re-training can be viewed as a special case of training.
As the re-training uses only legitimate data as training samples, the Trojans contained in the weights may be overwritten during the re-training process and hence the Trojans may be rendered inefficient.
Note that the re-training needs to be supervised, \ie the defender needs to know the label of each training sample.\par

Although re-training may use the same algorithm as training the neural IP from scratch, much fewer training samples are used in re-training.
This results in much faster convergence and hence the effort for re-training is much less than that of training the neural IP entirely in-house.\par

\subsection{Input Preprocessing} \label{Input_Pre}
The prior introduced defense approaches require some assumptions about neural IP and the defender.
The re-training approach requires the neural IP to be reconfigurable.
Both the re-training approach and the input anomaly detection approach require the defender to know the label of each legitimate sample.
These requirements are strong and are sometimes not satisfied.
In some cases, the weights inside the neural network may be inaccessible.
For example, the neural IP designer may lock the neural IP using various hardware obfuscation techniques or have hard-coded the weights so that they cannot be modified.
In some other cases, the defender may not necessarily know the label of each legitimate sample, \ie he/she indeed needs to rely on the neural IP for classification.
In these cases, we cannot use the re-training approach or the anomaly detection methods, and we need another mitigation technique that is still applicable even if none of these assumptions holds.
To this end, we propose an approach to preprocessing the input samples, \ie to insert an input preprocessor between the input and the neural IP.
The objective of input preprocessing is to prevent the illegitimate inputs from triggering the Trojans without affecting the classification accuracy of legitimate data.
\par

\begin{figure}[!t]
\centering
\includegraphics[width=0.2\textwidth]{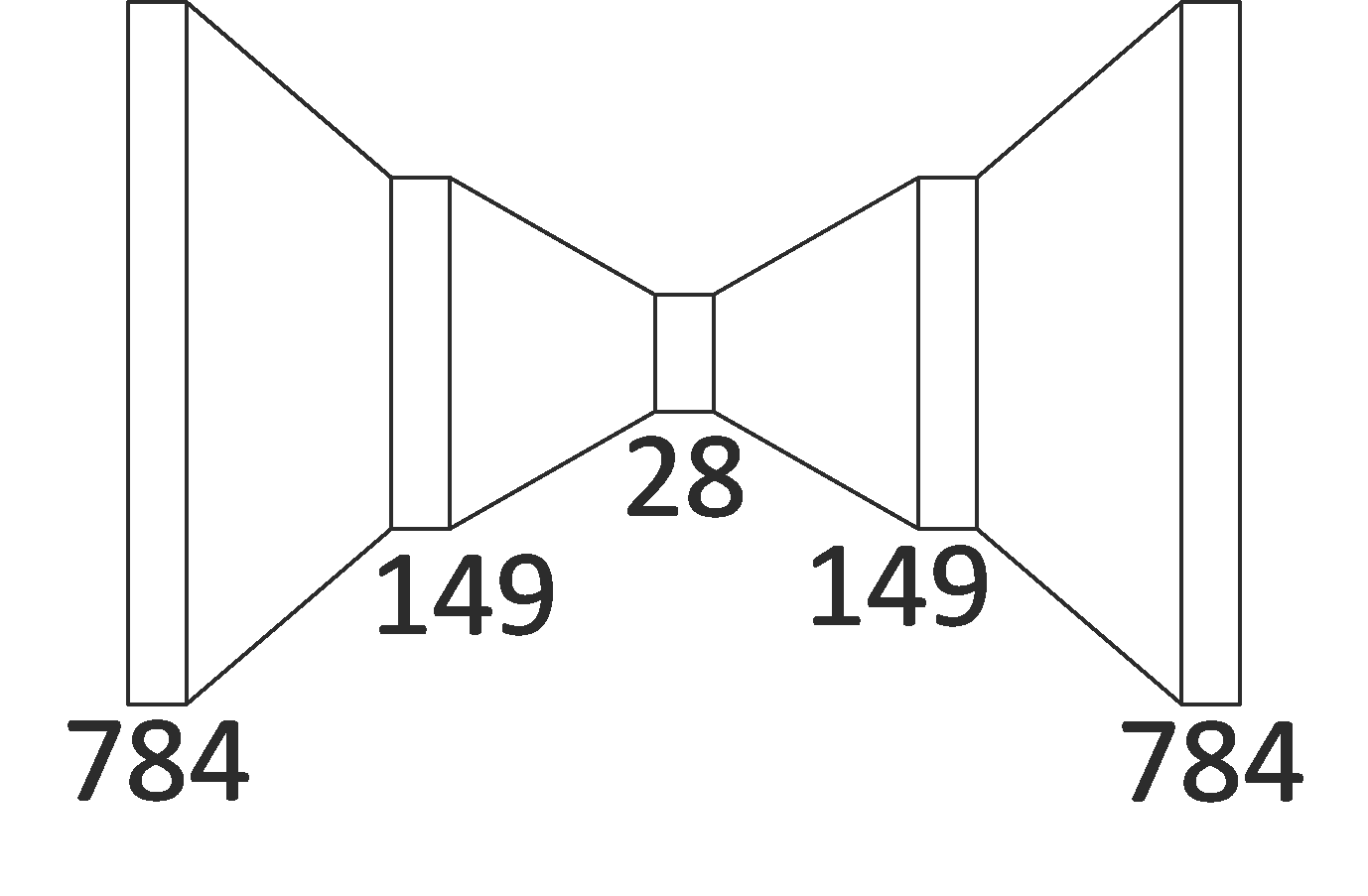}
\caption{The Architecture of the Autoencoder}
\label{autoenc_arch_fig}
\end{figure}
This objective reminds us of the autoencoder which is then used as the input preprocessor in this approach.
The autoencoder, a.k.a. the replicator neural network \cite{hawkins2002outlier}, is a neural network with the same number of input and output neurons and has a bottleneck structure.
The architecture of the autoencoder that we use in this work is shown in Fig. \ref{autoenc_arch_fig}.
In this figure, the rectangles stand for the neurons in each layer, the number beside each rectangle represents the number of neurons in the corresponding layer, and the trapezoids stand for the weights between adjacent layers.
The backpropagation algorithm is also used to train the autoencoder and the error function is given as
\begin{equation}
E(\mathbf{w},T) = \frac{1}{2n}\sum_{\mathbf{x}_i\in T}\|f(\mathbf{w},\mathbf{x}_i)-\mathbf{x}_i\|^2
\label{Error_Fn_Autoenc}
\end{equation}
From this error function, we can see that the objective of training an autoencoder is to minimize the mean square error between the images from the training set and the reconstructed images.
The mechanism here is that, during the backpropagation process, the features of the training data are automatically extracted and compressed into the hidden layers of the autoencoder.
Only legitimate data are used to train the autoencoder.
Therefore, during the test phase, it should be expected that if the input is from the legitimate distribution, the autoencoder's output should be close to the input,
and hence the neural IP should be able to classify the reconstructed image correctly as if it were the original input;
if the input is not from the illegitimate distribution, the reconstructed image should deviate a lot from the original input and hence should not be able to trigger the Trojans.
Note that, in this approach, no assumption is made about the neural IP and we only treat the neural IP as a black box.\par

\section{Experiments and Results} \label{Sec_Exp}
\subsection{Neural IP Setup}
In this work, the neural IP is supposed to classify the images of handwritten digits (from `0' to `9') from the MNIST dataset \cite{lecun1998gradient}.
We choose the neural network with 784 input neurons, 300 hidden neurons, and 10 output neurons as the architecture of the neural IP.
Each output neuron represents one possible classification result (\ie a label), and the label represented by the neuron which has the highest output value is the classification result.
60,000 legitimate samples (from the MNIST dataset) and 864 illegitimate samples are used in the training phase.
To ensure generality, as there are 10 different digits, we train 10 Trojan-embedded neural IP benchmarks and 1 Trojan-free benchmark.
For the $i^{th}$ Trojan-embedded benchmark, `$i-1$' is the label for the illegitimate data (e.g. the Trojan label is `0' for the first one and `9' for the last one, etc.).
The Trojan is said to be \textit{triggered} if an illegitimate input is classified as the Trojan label chosen by the attacker.
We define the \textit{Trojan activation rate} as the fraction of illegitimate input that triggers the Trojan.
The test dataset is composed of 10,000 legitimate samples and 152 illegitimate samples.
During test time, we observe the following results:
\begin{itemize}
\item The average Trojan activation rate is 99.2\% for the ten Trojan-embedded neural IPs.
\item For the Trojan-free neural IP, the classification accuracy (of legitimate samples) is 97.97\%, whereas the average classification accuracy of legitimate samples for the Trojan-embedded neural IPs is 97.77\%. 
\end{itemize}
These results show that the Trojans can be effectively triggered by the illegitimate inputs while the normal functionality of the neural IP is almost not affected by the Trojans.
Therefore, the threat of neural Trojans must be mitigated.

\subsection{Input Anomaly Detection}
\begin{table}[!t]
\centering
\caption{Anomaly Detection with Various Methods}
\begin{tabular}{|c  | c | c |}
\hline
Method &  Detection Rate & False Positive \\ \hline
SVM &  72.6\% & 13.4\% \\ \hline
Decision Tree & 99.8\% & 12.2\% \\ \hline
\end{tabular}
\label{Anom_Detec_Table}
\end{table}
As mentioned in Section \ref{Anom_Detec}, SVMs and DTs are implemented for input anomaly detection.
We train each SVM and DT with 60,000 legitimate samples.
The test data include 10,000 legitimate samples and 1016 illegitimate samples and the performance of each method is recorded in Table \ref{Anom_Detec_Table}.
The detection rate means the percentage of illegitimate inputs detected as anomalies, and the false positive means the percentage of legitimate inputs detected as anomalies.
Comparing the two approaches, we can find that the DTs achieve the better results: they are able to detect 99.8\% of the illegitimate inputs, although the false positive rate is 12.2\%.
Therefore, if there is a situation where a triggered Trojan can cost huge loss while a moderate false positive rate is acceptable, the DT-based anomaly detection approach should be desirable.\par

\subsection{Re-training}
Following the discussion in Section \ref{Re-training}, we re-train the neural IP benchmarks with legitimate data.
For each neural IP benchmark, we observe how the Trojan activation rate (for Trojan-embedded benchmarks only) and the classification accuracy of legitimate data change with the number of re-training samples.
We use up to 12,000 legitimate samples for re-training which is 20\% the number of legitimate samples used to train the neural IP.
As much fewer samples are used for re-training than training the neural IPs, the effort of re-training is substantially smaller than that of training a neural IP from scratch.\par

\begin{figure}
\centering
\includegraphics[width=0.45\textwidth]{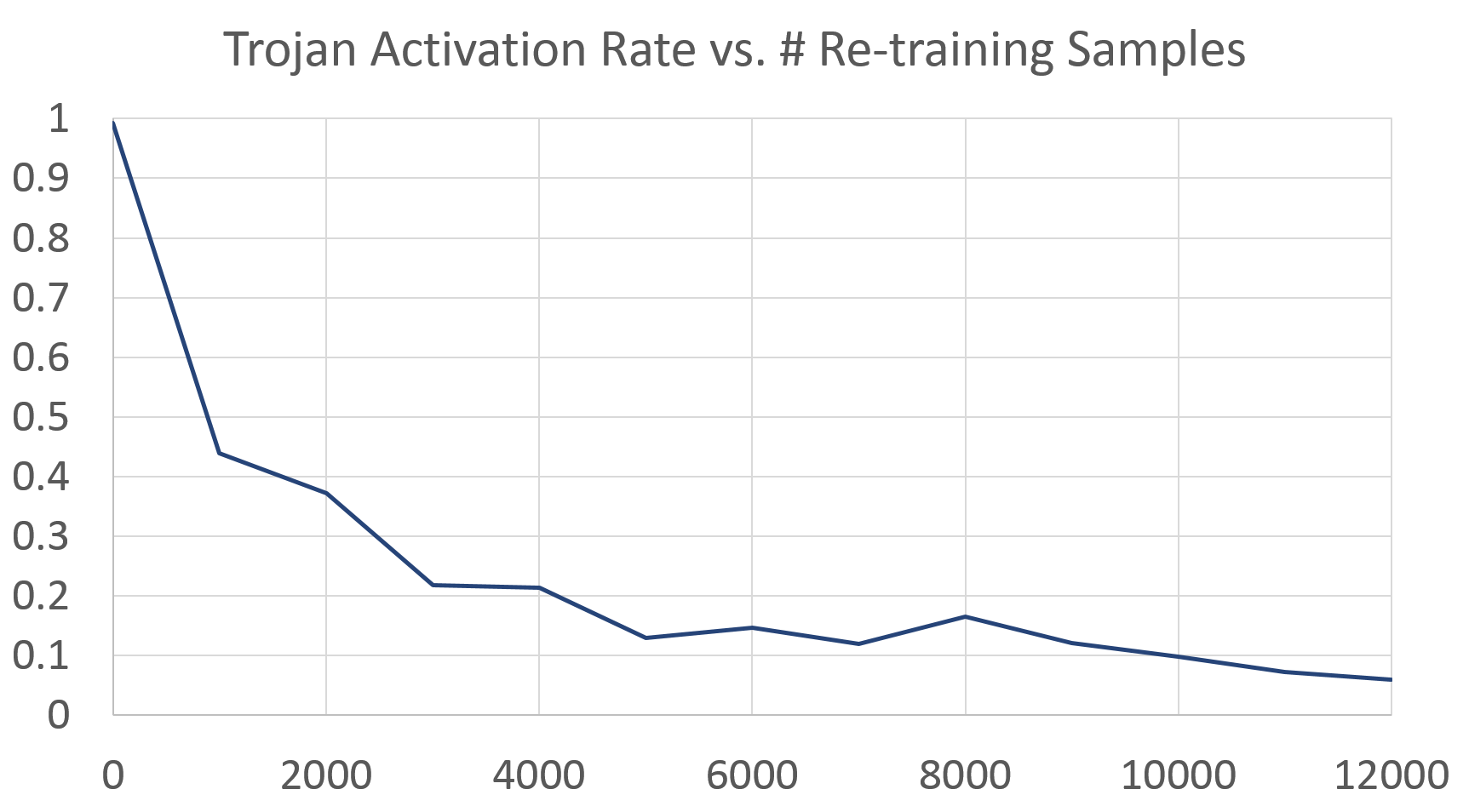}
\caption{The Average Trojan Activation Rate vs. Re-training Effort}
\label{TAR_Fig}
\end{figure}
The change of the average Trojan activation rate vs. the number of re-training samples over all the Trojan-embedded neural IP benchmarks is shown in Fig. \ref{TAR_Fig}.
It is shown that when the number of re-training samples exceeds 10,000, the Trojan activation rate drops below 10\% (5.9\% when 12,000 re-training samples are used).\par

\begin{figure}
\centering
\includegraphics[width=0.45\textwidth]{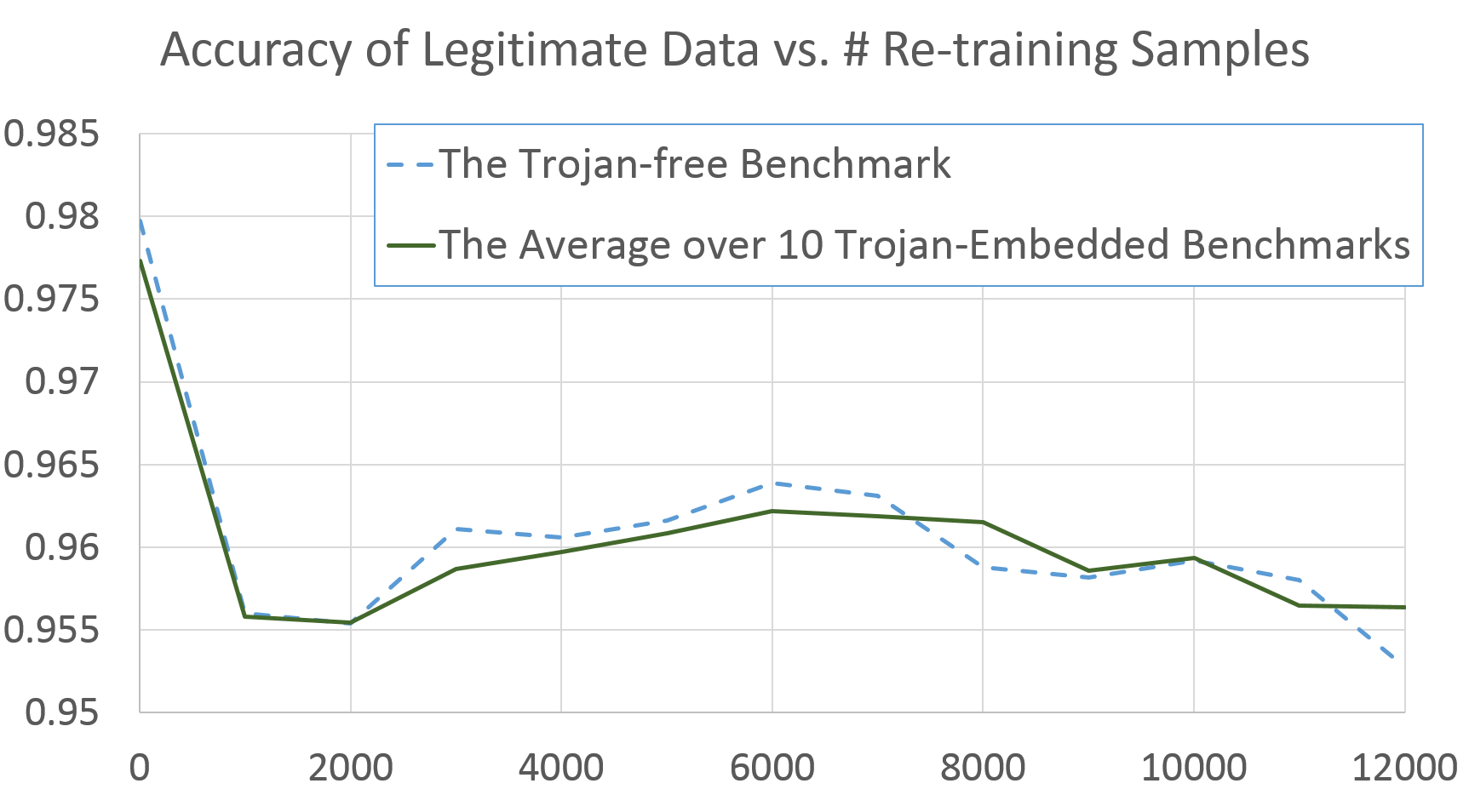}
\caption{The Average Classification Accuracy of Legitimate Data vs. Re-training Effort}
\label{Accu_Fig}
\end{figure}
The change of the classification accuracy of legitimate data vs. the number of re-training samples is shown in Fig. \ref{Accu_Fig}.
The dotted line stands for the Trojan-free neural IP and the solid line shows the average over all the Trojan-embedded benchmarks.
We observe that the re-training results in a decrease of about 2\% in the classification accuracy of legitimate data for both the Trojan-free and the Trojan-embedded benchmarks.
A possible reason is that we are only using a small subset of legitimate samples and they do not represent the distribution of the legitimate samples very well.\par

In summary, the re-training approach proves effective in reducing the Trojan activation rate and it requires substantially less effort than training a neural IP in-house. However, it suffers from the following limitations:
\begin{itemize}
\item The neural IP will suffer from an average of 2\% reduction in the classification accuracy of legitimate data no matter the neural IP is Trojan-embedded or Trojan-free.
\item There are strong assumptions about the neural IP and the defender: the neural IP must be re-trainable and the defender must know the label of each legitimate sample.
\end{itemize}

\subsection{ Input Preprocessing}
\begin{figure}[!t]
\centering
\subfloat[The Original (Upper Row) and Reconstructed (Lower Row) Legitimate Input Images ]{\includegraphics[width=0.45\textwidth]{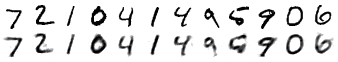}\label{recons_orig_fig}}
\hfill
\subfloat[The Original (Upper Row) and Reconstructed (Lower Row) Illegitimate Input Images ]{\includegraphics[width=0.45\textwidth]{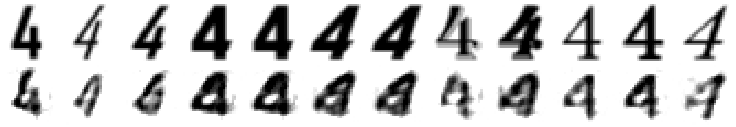}\label{recons_trojan_fig}}
\caption{The Original and Reconstructed (a) Legitimate and (b) Illegitimate Input Images}
\label{Recons_Fig}
\end{figure}

As mentioned in Section \ref{Input_Pre}, we use an autoencoder for input preprocessing.
The autoencoder we use in this work has 3 hidden layers.
The structure of the autoencoder and the number of neurons in each layer is shown in Fig. \ref{autoenc_arch_fig} where the rectangles stand for the neurons in each layer and the trapezoids stand for the weights between adjacent layers.
The logistic sigmoid function, \ie $y = \frac{1}{1+e^{-x}}$, is used as the activation function of the middle layer, and the ReLU function is the activation function of all other layers.\par

The autoencoder is trained with 60,000 legitimate samples and we test this approach with 1016 illegitimate input samples and 10000 legitimate input samples.
Fig. \ref{Recons_Fig} demonstrates how the autoencoder reconstructs the input images.
In Fig. \ref{recons_orig_fig}, the upper row contains some samples of legitimate input images, and the corresponding reconstructed images by the autoencoder is shown in the lower row.
It can be observed that the reconstructed images are similar to the actual input images.
Therefore, that the neural IP should be able to classify the reconstructed images correctly as if they were the original input.
On the other hand, if the input images are illegitimate, as shown in the upper row of Fig. \ref{recons_trojan_fig}, the reconstructed images (illustrated in the lower row) will suffer from much larger distortion.
In some cases, it is not even clear to a human observer which digit the reconstructed image is of.
Therefore, the neural IP should not be able to recognize the reconstructed illegitimate images as Trojan triggers.\par

Our experiments show that 90.2\% of the Trojan triggers are disabled in this approach.
Furthermore, we found that, with the input preprocessor in place, the behavior of the Trojan-embedded neural IPs is very similar to that of the Trojan-free neural IP:
 in 96.8\% of the cases where the illegitimate inputs are given, the outputs of the Trojan-embedded neural IPs are the same as the outputs of the Trojan-free neural IP.
Therefore, the Trojans are rendered useless.
The impact of input preprocessing on the classification accuracy of legitimate data is rather small:
for the Trojan-free benchmark,  this accuracy is 96.97\%, 1.00\% lower than that without the input preprocessor;
for the 10 Trojan-embedded neural IPs, the average accuracy is 95.41\%, 2.36\% lower than that without using the autoencoder.\par


\section{Summary and Conclusion} \label{Sec_Conc}
In this work,  we first reviewed the existing security threats to neural networks, including the poisoning attack and the exploratory attack.
In these attack models, the attacker wants to either weaken the network by providing poisoned training samples or find the vulnerabilities of the network by generating adversarial test samples.\par

In addition to these attack scenarios, we propose the neural Trojan attack which is carried out by the trainer of the neural IP.
The attacker can train the neural IP to recognize a certain illegitimate pattern (\ie the Trojan trigger) and produce an output in favor of the attacker in addition to training the neural IP to have the normal functionality.
We have shown that the attacker can indeed train the neural IP so that the Trojan is triggered in more than 99\% of the cases where the Trojan trigger is given without significantly affecting the normal functionality of the neural IP.\par

The defender is the system designer who buys the neural IP from the attacker, but he/she does not know whether there is a Trojan embedded into the neural IP.
To mitigate this threat, we propose three techniques: input anomaly detection, re-training, and input preprocessing.
The input anomaly detection approach is able to detect 99.8\% of the illegitimate input samples but at the price of 12.2\% false positive.
The re-training approach can reduce the Trojan activation rate to 6\% and the effort for re-training is substantially less than training the neural IP in-house.
However, this approach needs the neural IP to be re-trainable.
In the input pre-processing approach, we reconstruct the input image with an autoencoder and use the reconstructed image as the input to the neural IP.
In this way, we are able to disable 90.2\% of the Trojan triggers without requiring any knowledge about the neural IP.\par

In conclusion, we have demonstrated that the threat from neural Trojans must be considered when we use a neural IP obtained from elsewhere, and we have also proposed three countermeasures that system designer can choose from when using such a neural IP that potentially contains Trojans.
Although all these approaches are proven effective in mitigating the threat of neural Trojans, they all come with some overheads: the reduction in the accuracy of legitimate data, the rejection of some legitimate inputs, etc.
Therefore, our future work may include finding mitigation approaches that are more effective and have lower overheads.


\bibliographystyle{IEEEtran}
\bibliography{IEEEabrv,reference}

\end{document}